\documentclass[aps,prd,twocolumn,showpacs,10pt,superscriptaddress,preprintnumbers,nofootinbib,floatfix]{revtex4-2}
\usepackage{amsmath}
\usepackage{amssymb}
\usepackage{bm}
\usepackage{slashed}
\usepackage{graphicx}
\usepackage{color}
\begin{document}

\title{Angular-Momentum Detector Operators in Massless Yukawa Theory}

\author{Mian Fu}
\email{mianfu25@stu.pku.edu.cn}
\affiliation{School of Physics, Peking University, Beijing 100871, China}
\author{Hua Xing Zhu}
\email{zhuhx@pku.edu.cn}
\affiliation{School of Physics, Peking University, Beijing 100871, China}
\affiliation{Center for High Energy Physics, Peking University, Beijing 100871, China}
\affiliation{Southern Center for Nuclear-Science Theory (SCNT), Institute of Modern Physics,
Chinese Academy of Sciences, Huizhou 516000, Guangdong Province, China}
\author{Hong-Hao Zhuang}
\email{alvinzhuang@stu.pku.edu.cn}
\affiliation{School of Physics, Peking University, Beijing 100871, China}

\begin{abstract}
Angular-momentum flux with
directional resolution at null infinity can provide a new experimental probe of quantum field theory compared with energy flux. We construct angular-momentum detector operators for massless scalar and Dirac fields, which in principle can be realized as a detector in particle collisions. For a conserved vector-current source in massless Yukawa theory, we compute the complete one-loop matrix element for angular-momentum detector. We find that for this conserved-current correlator in massless Yukawa theory, the Belinfante flux is infrared finite through one loop. The canonical flux would define a different angular-momentum detector, because the moment arm makes the subleading momentum flux contribute at null infinity. Both the canonical correlator and the correlator obtained by isolating the spin terms in the Belinfante detector retain a collinear pole at one loop. We illustrate the use of angular-momentum detector with different polarized wave packets. We show that the angular-momentum detector is able to probe first momentum derivatives of the density-matrix kernel at coincident total momenta, which is inaccessible to the usual energy operator.

\end{abstract}

\maketitle

\section{Introduction}

Conservation laws and conserved charges have been central to physics
throughout its history. In modern particle physics, energy, the most
basic conserved quantity, has been studied extensively. At colliders,
energy-momentum measurements have developed into a mature field, and
energy-energy correlators are widely used in theoretical and experimental
studies to probe the dynamics of quantum field theory~\cite{Basham:1977iq,Basham:1978bw,Basham:1978zq}, see \cite{Moult:2025nhu} for a review. Recently there has also been significant experimental progress in measuring energy correlators at various colliders, revealing the rich structure of Quantum Chromodynamics~(QCD) and the transition from partonic dynamics to hadronization~\cite{Komiske:2022enw,CMS:2024mlf,ALICE:2024dfl,ALICE:2025igw,Tamis:2023guc,STAR:2025jut,CMS:2024ovv,CMS:2025ydi,Bossi:2025nux}.

A natural theoretical framework for describing these direction-resolved measurements is provided by detector operators, also known as light-ray operators~\cite{Sveshnikov:1995vi,Korchemsky:1999kt,Hofman:2008ar,Kravchuk:2018htv,Caron-Huot:2022eqs}. Defined at null infinity, these operators encode the fluxes of conserved quantities along specified directions and thus provide an operator-level description of observables such as energy correlators.
They have also been
studied extensively in the context of
the averaged and quantum null energy conditions
\cite{Faulkner:2016mzt,Balakrishnan:2017bjg}, modular Hamiltonians and
quantum-information methods \cite{Casini:2017roe}, causality
constraints \cite{Hartman:2016lgu}, asymptotic symmetries
\cite{Bondi:1962px,Sachs:1962wk,Sachs:1962zza,Cordova:2018ygx}, and
soft graviton theorems \cite{Strominger:2013jfa,He:2014laa}. Within
conformal field theory, the light transform provides a systematic
construction of nonlocal light-ray operators \cite{Kravchuk:2018htv},
and the algebra of stress-tensor light-ray operators, including
coordinate moments, has been studied in conformal field theory
\cite{Cordova:2018ygx,Belin_2021}. Related hard-mode detector algebras
have also been investigated for four-dimensional theories of massless
integer-spin fields \cite{Gonzo_2021}. 
These developments establish detector operators as a general framework for studying conserved-quantity fluxes at null infinity. While energy-flow observables have been extensively investigated, their angular-momentum counterparts raise additional questions, particularly concerning their collinear behavior and infrared safety.

Energy weights suppress soft radiation and are additive under collinear splitting, making energy correlators infrared safe. These properties ensure that the measurement is insensitive to unresolved soft emissions and collinear splittings, allowing real--virtual singularities to cancel according to the Kinoshita--Lee--Nauenberg theorem~\cite{Kinoshita:1962ur,Lee:1964is}. This infrared safety underlies higher-order perturbative calculations and applications of energy correlators to jet substructure~\cite{Belitsky:2013ofa,Dixon:2018qgp,Henn:2019gkr,Dixon:2019uzg,Korchemsky:2019nzm,Kologlu:2019mfz,Chen:2020vvp}. For more general weighted scattering observables, however, symmetry and conservation laws alone do not guarantee infrared safety; the measurement must also remain insensitive to unresolved emissions. Angular-momentum flux provides an interesting case in which this requirement is nontrivial, since its detector operators involve momentum derivatives that can modify the collinear singularity structure. Consequently, unlike energy-flow observables, the infrared behavior of direction-resolved angular-momentum correlators requires a separate investigation. In this work, we study the collinear singularity structure of angular-momentum flux detectors in massless Yukawa theory. In particular, we investigate their dependence on the choice between canonical and Belinfante energy-momentum tensors, as well as their sensitivity to first momentum derivatives of the density-matrix kernel at coincident total momenta.

In QCD, decomposing the nucleon spin into quark and gluon contributions
has long been a challenge, in part because gauge-invariant and physically
meaningful separations of spin and orbital angular momentum are difficult
to define \cite{Jaffe:1989jz,Ji:1996ek,Leader:2013jra}. Related issues
also arise in perturbative QED, where a complete one-loop calculation
yields equal electron orbital angular momenta in the Jaffe--Manohar and
Ji decompositions \cite{Ji:2015sio}.
Understanding the angular momentum decomposition is an important scientific goal of the upcoming Electron-Ion Collider, see e.g.~\cite{Ji:2016jgn,Ji:2020ena,Hatta:2016aoc,Kovchegov:2017lsr,Boussarie:2019icw,Bhattacharya:2022vvo,Cougoulic:2022gbk,Bhattacharya:2023hbq,Bhattacharya:2024sno,Bhattacharya:2024sck}.
Rather than study the nucleon's
internal spin structure, as encoded in the spatial
distributions of angular momentum \cite{Lorce:2017spatial}, we focus on
the angular-momentum flux of massless particles produced in scattering at null infinity. Measuring the flux over the
celestial sphere may reveal features of the source that are inaccessible
in the usual energy and momentum measurement, and might provide useful implication to the nucleon spin-decomposition
program. 

It is well-known that spin information can be inferred from decay angular distributions, and has enabled
observations of entanglement in top-quark pairs
\cite{ATLAS:2024Entanglement,CMS:2024Entanglement},
Bell-like tests in hyperon systems
\cite{BESIII:2025Bell,BESIII:2026Spin},
and an experimental characterization of a broader hierarchy of quantum correlations
\cite{Afik:2026TopCorrelations}.
These analyses reconstruct spin information from decay products,
whereas a direction-resolved angular-momentum flux is a different
observable whose detector interpretation is developed here. Although direction-resolved angular-momentum measurements have
not yet been realized at colliders, the detector correlators studied here
provide a theoretical foundation for possible future experimental investigations.

The structure  of the paper is as follows. In Sec.~\ref{sec:defs}, we define the angular-momentum operator. In Sec.~\ref{sec:corr}, we introduce the conserved vector current operator into the correlator and discuss some properties caused by spatial rotation and global $U(1)$ symmetry. Then, in Sec.~\ref{sec:oneloop}, we present the complete infrared-safe result of the correlator with Belinfante's flux definition up to NLO in Yukawa theory. To demonstrate the angular-momentum detection in Yukawa theory more intuitively, we display the results with wave packets in Sec.~\ref{sec:example}. 

\section{Definitions and operator actions\label{sec:defs}}

We work in $D$-dimensional Minkowski space with the
metric $g^{\mu\nu}=\mathrm{diag}(+,-,\ldots,-)$. The detector
direction is the null vector $n^{\mu}=(1,\hat n)$, with
$\bm{x}=r\hat n$ and retarded time $u=t-r$; $\theta$ and
$\varphi$ denote the polar and azimuthal angles of $\hat n$.
Throughout, $n^{2}$ and
$q^{2}$ denote the Lorentz norms $n_{\mu}n^{\mu}=0$ and
$q_{\mu}q^{\mu}$.

The Belinfante angular-momentum current~\cite{Belinfante:1939oxq} associated with rotations in the
$a-b$ plane is
\begin{equation}
  I^{\mu ab} = x^{a} \Theta^{\mu b} - x^{b} \Theta^{\mu a},\label{eq:local_I}
\end{equation}
where the symmetric energy-momentum tensor $\Theta^{\mu\nu}$ is
obtained from the canonical tensor $T^{\mu\nu}$ through the Belinfante improvement
\begin{equation}
  \Theta^{\mu\nu} = T^{\mu\nu}
  + \frac12\partial_{\lambda}\left(\mathcal{S}^{\lambda\mu\nu}
  + \mathcal{S}^{\mu\nu\lambda} + \mathcal{S}^{\nu\mu\lambda}\right),
\end{equation}
and $\mathcal{S}^{\lambda\mu\nu} = (-\mathrm i \Sigma^{\mu\nu})_{ij} \psi_j \frac{\delta \mathcal L}{\delta(\partial_\lambda \psi_i)}$ is the spin current for arbitrary fields $\psi$ involved in Lagrangian $\mathcal{L}$ with the spin matrix $\Sigma^{\mu\nu}$, where we use the left functional derivative.
The direction-resolved fluxes are measured by the detector operators
\begin{equation}
  \begin{aligned}
    \mathcal{E}(\hat n)
    &= \lim_{r\to+\infty} r^{D-2}
    \int_{0}^{+\infty}\mathrm{d}t\, n^i\, \Theta^{0i}(t,r\hat n),\\
    \mathcal{M}^{ab}(\hat n)
    &= \lim_{r\to+\infty} r^{D-2}
    \int_{0}^{+\infty}\mathrm{d}t\, n^i\, I^{iab}(t,r\hat n),
  \end{aligned}
  \label{eq:lightray-M}
\end{equation}
the first being the total energy flux, summed over all particle species,
which enters the averaged null energy
condition~\cite{Faulkner:2016mzt,Hartman:2016lgu,Kravchuk:2018htv}, and
the second the angular-momentum flux.

For a massless real scalar, whose canonical energy-momentum tensor is
symmetric, define the radiation field
$\Phi(u,\hat n)\equiv\lim_{r\to\infty}r^{(D-2)/2}
\phi(t=r+u,r\hat n)$ and the notation of infinitesimal rotations
$\Lambda^{ab}A \equiv \mathrm{i}(n^a\nabla^b-n^b\nabla^a)A$
and the spherical derivatives
$\nabla^a=-(\delta^{ab}-n^a n^b)\partial/\partial n^b$,
the scalar angular-momentum detector operator \eqref{eq:lightray-M} takes the
radiation-field form
\begin{equation}
  \mathcal{M}_S^{ab}(\hat n)
  = -\mathrm{i}\int_{-\infty}^{+\infty}\mathrm{d}u\,:
  \partial_u\Phi\,\Lambda^{ab}\Phi:,
  \label{eq:M-scalar}
\end{equation}
$\Lambda^{ab}$ acts on the angular dependence of $\Phi$. For a real
scalar, the canonical and Belinfante forms coincide, and
integrating Eq.~\eqref{eq:M-scalar} over the full sphere gives the total
angular-momentum charge $M^{ab}$.

Using covariantly normalized plane-wave modes,
$[c_{\bm k},c^\dagger_{\bm p}]
=2k^0(2\pi)^{D-1}\delta^{(D-1)}(\bm k-\bm p)$,
the same operator becomes
\begin{equation}
\begin{aligned}
  \mathcal{M}_S^{ab}(\hat n)
  ={}\frac{1}{4(2\pi)^{D-1}}
  \int_0^{+\infty}\mathrm{d}\omega\,\omega^{D-3}
  c^\dagger_{\omega\hat n}\overleftrightarrow
  \Lambda^{ab}c_{\omega\hat n}.
\end{aligned}
\label{eq:M-scalar-modes}
\end{equation}

For the massless Dirac fermions, the radiation field
is $\Psi(u,\hat n)\equiv\lim_{r\to\infty}r^{(D-2)/2}\psi$.
The Belinfante angular-momentum detector operator is
\begin{equation}
\begin{aligned}
  \mathcal{M}_{D}^{ab}(\hat n)
  &= \frac12\int^{+\infty}_{-\infty}\mathrm{d}u\, \Bigl(:\Psi^{\dagger}
  \overleftrightarrow{\Lambda}^{ab}\Psi: \\
  & - n^a \nabla_{c}:\Psi^\dagger \Sigma^{cb} \Psi: + n^b \nabla_{c}:\Psi^\dagger \Sigma^{ca} \Psi:\Bigr),
\end{aligned}\label{eq:M-Dirac}
\end{equation}
where $\Psi^\dagger\overleftrightarrow{\Lambda}^{ab}\Psi=
\Psi^\dagger(\Lambda^{ab}\Psi)
-(\Lambda^{ab}\Psi^{\dagger})\Psi$ and the spin matrices 
$\Sigma^{ab} = \frac{\mathrm{i}}{4}[\gamma^a,\gamma^b]$.

Using the same covariant normalization for the fermion
anticommutators and
$\chi^{r\dagger}_{\bm k}\chi^s_{\bm k}=2\omega\delta^{rs}$,
we choose a local polarization frame adapted to the fixed rotation plane,
$(\Lambda^{ab}+\Sigma^{ab})\chi^s_{\bm k}=0$.
The plane-wave mode representation is then
\begin{equation}
\begin{aligned}
  \mathcal{M}_D^{ab}(\hat n)
  ={}&\frac{1}{4(2\pi)^{D-1}}
  \int_0^{+\infty}\mathrm{d}\omega\,\omega^{D-3}\\
  &\times\Biggl[
  \sum_s\Bigl(
  a_{\omega\hat n}^{s\dagger}\overleftrightarrow{\Lambda}^{ab}a_{\omega\hat n}^s
  +b_{\omega\hat n}^{s\dagger}\overleftrightarrow{\Lambda}^{ab}b_{\omega\hat n}^s
  \Bigr)\\
  &\qquad-\nabla_c
  \Bigl(n^a \sigma_{\omega\hat n}^{cb}-n^b \sigma_{\omega\hat n}^{ca}\Bigr)
  \Biggr],
\end{aligned}
\label{eq:M-Dirac-modes}
\end{equation}
where all mode operators are evaluated at $\bm k=\omega\hat n$, and
\begin{equation}
\begin{aligned}
 \sigma_{\omega\hat n}^{ab}
 &=\sum_{r,s} \frac{\chi_{\omega \hat n}^{r\dagger}\Sigma^{ab}\chi_{\omega \hat n}^s}
 {2\omega} \Bigl(
 a_{\omega\hat n}^{r\dagger}a_{\omega\hat n}^s
 -b_{\omega\hat n}^{s\dagger}b_{\omega\hat n}^r
 \Bigr).
\end{aligned}
\label{eq:spin-density-modes}
\end{equation}

Actually, different definitions of local angular-momentum flux may produce different detector operators defined in Eq.~\eqref{eq:lightray-M}, among which the detector of canonical definition is discussed in Appendix~\ref{app:canonical}. Difference between local flux definitions can be expressed as a zero-divergence total derivative term, which does not affect the energy detector for either the scalar or the Dirac field. For the Dirac field the leading radial flux cancels in the antisymmetric projection. The finite-radius Dirac equation expresses the surviving flux in terms of angular derivatives of the radiation field, as derived in Appendix~\ref{app:subleading}. In 4-dimensional space-time, the detector can also be derived with the spherical-wave modes as is shown in Appendix~\ref{app:sperical}.

\section{Distributional correlators and conservation law\label{sec:corr}}

We study
$C^{ab;\mu\nu} \equiv \langle\tilde J^{\mu}(q)\,\mathcal{M}^{ab}(\hat n)\,
\tilde J^{\nu\dagger}(q')\rangle$, where
$J^{\mu}=e\mu_R^\epsilon:\bar\psi\gamma^{\mu}\psi:$ in dimensional 
regularization $D = 4 - 2\epsilon$ is the conserved vector current of a source of
charge $e$, and
$\tilde J^{\mu}(q)=\int\mathrm{d}^{D}x\,J^{\mu}(x)\,
\mathrm{e}^{\mathrm{i}q\cdot x}$. We restrict to future-timelike
$q$, suppressing the common support
$\theta(q^0)\theta(q^2)$. 
Spinor traces use $N_{D}\equiv\operatorname{tr}\bm{1}$, fixed
at $N_D=4$ in dimensional regularization. 

To tackle this Wightman correlator, we here insert two identity operators, one on each side of the detector, changing it into a form of weighted total cross-section,
\begin{equation}
  \small
  C^{ab;\mu\nu} = \sum_{X,Y} \langle 0 | \tilde J^{\mu}(q)| X \rangle \langle X | \mathcal M^{ab}(\hat n) | Y \rangle \langle Y | \tilde J^{\nu \dagger} (q') | 0 \rangle, 
\end{equation}
where $X$ and $Y$ include all possible final states in the phase space, carrying total momentum $q$ and $q'$, respectively. One should notice that it is necessary to assume the independence between momenta $q$ and $q'$ to achieve the generally correct result.

In the Poincaré group, rotations act differentially in momentum space
and do not commute with translations. Consequently, the total
momentum-conservation factor
$\Delta_{qq'}\equiv(2\pi)^{D}\delta^{D}(q-q')$ cannot in general be
factored out of the correlator. The correlator therefore contains
derivatives of the momentum-conserving delta function, which we
abbreviate as
\begin{equation}
  \mathcal{D}^{ab} \Delta_{qq'}\equiv (2\pi)^{D} \mathrm{i}
  \left(n^{a}\partial_{q}^{b}-n^{b}\partial_{q}^{a}\right) \delta^{D}(q-q'),
  \label{eq:Ddef}
\end{equation}
with $\partial_{q}^{b}\equiv\partial/\partial q_{b}$ denoting
differentiation with respect to a covariant momentum component. The
complete correlator can therefore be decomposed as
\begin{equation}
  C^{ab;\mu\nu}
  = F^{ab;\mu\nu}(q,n)\,\Delta_{qq'}
  + G^{\mu\nu}(q,n)\,\mathcal{D}^{ab}\Delta_{qq'}.
  \label{eq:Cform}
\end{equation}

Ward identities hold in the distributional sense on both
sources. The product rules
\begin{equation}
\begin{aligned}
q'_\nu\mathcal D^{ab}\Delta_{qq'}
&=q_\nu\mathcal D^{ab}\Delta_{qq'}+\mathrm{i}{\kappa^{ab}}_\nu\Delta_{qq'},\\
\kappa^{ab;\mu}&\equiv n^ag^{b\mu}-n^bg^{a\mu}.
\end{aligned}
\label{eq:distid}
\end{equation}
together with $q_{\mu}G^{\mu\nu}=G^{\mu\nu}q_{\nu}=0$ reduce
the two Ward identities to
\begin{equation}
  q_{\mu}F^{ab;\mu\nu}=0,
  \qquad
  F^{ab;\mu\nu}q_{\nu}=-\mathrm{i}G^{\mu\nu}{\kappa^{ab}}_{\nu},
  \label{eq:ward0}
\end{equation}
both satisfied by the Belinfante free correlators.

Canonical quantization of the total charge
$M^{ab}=\int\mathrm{d}^{D-1}x\,I^{0ab}$ gives
\begin{equation}
  [M^{ab},J^{\mu}(y)]
  = -(g^{\mu\nu}\Lambda_{y}^{ab}
  +\mathrm{i}g^{\mu a}g^{\nu b}
  -\mathrm{i}g^{\mu b}g^{\nu a})J_{\nu}(y),
  \label{eq:etcr}
\end{equation}
with $\Lambda_{y}^{ab}=\mathrm{i}(y^{a}\partial_{y}^{b}
-y^{b}\partial_{y}^{a})$; the local density $I^{0ab}(x)$ itself
carries additional derivative-of-delta contact terms that drop out of
the spatial integral. Since $M^{ab}$ annihilates the rotationally
invariant vacuum, Gauss's theorem fixes the integrated correlator at
all orders in perturbation theory,
\begin{equation}
\begin{aligned}
\int \mathrm{d}\Omega_{\hat n}\,C^{ab;\mu\nu}
={}\Lambda_q^{ab}\Pi^{\mu\nu}+\mathrm{i}(g^{\mu a}g^{\rho b}-g^{\mu b}g^{\rho a})
\Pi_\rho{}^\nu,
\end{aligned}
\label{eq:conservation}
\end{equation}
where $\Pi^{\mu\nu}\equiv
\langle\tilde J^\mu(q)\tilde J^{\nu\dagger}(q')\rangle$. This total angular-momentum sum rule
imposes a nontrivial constraint on the coefficients of these various
tensors, but it constrains only their integrated value.

\section{Correlator in Yukawa theory to NLO\label{sec:oneloop}}

To illustrate how the angular-momentum detector operator acts on a
correlator, we consider massless Yukawa theory with interaction
$\mathcal{L}_{\mathrm{int}}=g_{c}\mu_R^\epsilon\bar\psi\psi\phi$, whose discrete
symmetry is discussed in Appendix~\ref{app:symmetry}. We use
$s\equiv q^{2}$, $w\equiv n\cdot q$, and the
minimal-subtraction (MS) scale $\mu_{R}$. Contracting the correlator with two
polarization vectors describes the decay $\gamma^*\to\psi\bar{\psi}(\phi)$.

At tree level the coefficients are
\begin{gather}
  \begin{aligned}
    &F^{(0)ab;\mu\nu}
    = \frac{\mathrm{i}e^2\mu_R^{2\epsilon}N_{D}s^{D-2}}{2^{D}(2\pi)^{D-2}w^{D}}\,\biggl[\\
  &\quad -A^{ab}\Bigl((D-1)w\,P^{\mu\nu}
        + \frac{s}{w}n^{\mu}n^{\nu}\Bigr)\\
    &\quad+ \frac{s}{2}W_{+}^{ab\mu\nu}(n,n)
    + \frac{D-2}{2}s\,W_{-}^{ab\mu\nu}(n,n)\\
    &\quad- \frac{D-1}{2}w\,W_{-}^{ab\mu\nu}(n,q)
  - \frac{D-1}{2}A^{ab}A^{\mu\nu}\biggr],
  \end{aligned}\label{eq:F0}\\
  G^{(0)\mu\nu} = - e^2 \mu_R^{2\epsilon}
    \frac{N_{D}s^{D-1}}{2^{D}(2\pi)^{D-2}w^{D-1}}
    P^{\mu\nu}\label{eq:G0},
\end{gather}
with the following tensors, where the subscripts $+$ and $-$ indicate
symmetry and antisymmetry under $\mu\leftrightarrow\nu$, respectively:
\begin{gather}
  A^{\rho\sigma} = n^{\rho}q^{\sigma}-n^{\sigma}q^{\rho},
  \label{eq:A}\\
  H^{\mu\nu} = q^{\mu}q^{\nu}-g^{\mu\nu}s,
  \label{eq:H}\\
  P^{\mu\nu} = g^{\mu\nu}-\frac{q^{\mu}n^{\nu}+n^{\mu}q^{\nu}}{w}
  +\frac{s}{w^{2}}n^{\mu}n^{\nu},
  \label{eq:AP}\\
  \begin{aligned}
    W_{\pm}^{ab\mu\nu}(x,y) &= y^{\mu}\left(x^{a}g^{b\nu}
    -x^{b}g^{a\nu}\right)\\
    &\quad \pm y^{\nu}\left(x^{a}g^{b\mu}
    -x^{b}g^{a\mu}\right).
  \end{aligned}
  \label{eq:W}
\end{gather}

\begin{figure}
  \includegraphics[width=0.45\textwidth]{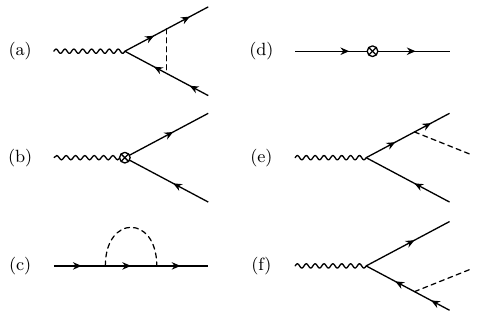}
  \caption{\label{fig:oneloop}One-loop corrections to the current
  correlator in the Yukawa
  theory: (a) one-loop vertex correction to the electromagnetic current;
  (b) ultraviolet counterterm for the vertex; (c) one-loop correction to
  the on-shell fermion propagator; (d) ultraviolet counterterm for the
  propagator; (e), (f) real emission of a Yukawa scalar.}
\end{figure}

Including the Yukawa interaction adds virtual and real corrections, shown
in Fig.~\ref{fig:oneloop}.
The coefficients $F$ and $G$ expanded at $D=4$ are
\begin{equation}
    \begin{aligned}
    &F^{(1)ab;\mu\nu}
    = \frac{\mathrm{i}e^2g_c^2s^2}{256\pi^4w^4}\,\biggl[\\
  &\quad A^{ab}\Bigl(9w\,P^{\mu\nu}
      + \frac{3s}{w}n^{\mu}n^{\nu}
    + \frac{2w}{s}H^{\mu\nu} - w\,g^{\mu\nu}\Bigr)\\
  &\quad {}- \frac{3s}{2}W_{+}^{ab\mu\nu}(n,n)
    + \frac{w}{2}W_{+}^{ab\mu\nu}(n,q)
    + 3A^{ab}A^{\mu\nu}\\
  &\quad {}- \frac{3s}{2}W_{-}^{ab\mu\nu}(n,n)
    + \frac{5w}{2}W_{-}^{ab\mu\nu}(n,q)\biggr],
    \end{aligned}\label{eq:F1}
\end{equation}
\begin{equation}
  G^{(1)\mu\nu}
  = \frac{e^2g_{c}^{2}s^{3}}{256\pi^{4}w^{3}}
    \left(3P^{\mu\nu}+\frac{H^{\mu\nu}}{s}\right),\label{eq:G1}
\end{equation}
All $1/\epsilon$ poles and scale logarithms cancel in each tensor
coefficient, so the complete one-loop correlator is finite without a
detector counterterm. Through one loop, it takes the compact form
\begin{equation}
  C^{(0+1)}
  = \left(1-\frac{g_{c}^{2}}{8\pi^{2}}\right)C^{(0)}
  +\frac{g_{c}^{2}}{16\pi^{2}}\widehat C+O(g_{c}^{4}),
  \label{eq:C01}
\end{equation}
where $C^{(0)ab;\mu\nu}$ is the free correlator and
\begin{equation}
\begin{aligned}
&\widehat C^{ab;\mu\nu}
=\frac{e^2s^2}{16\pi^2w^3}\left(sP^{\mu\nu}+H^{\mu\nu}\right)\mathcal D^{ab}\Delta_{qq'}\\
&\quad\;+\frac{\mathrm{i}e^2s^2}{16\pi^2w^3}\left[A^{ab}R^{\mu\nu}+\left(q^\nu-\frac{s}wn^\nu\right)\kappa^{ab;\mu}\right]\Delta_{qq'},
\end{aligned}
\label{eq:Chat}
\end{equation}
with
\begin{equation}
R^{\mu\nu}=\frac{4s}{w^2}n^\mu n^\nu
-\frac3w(n^\mu q^\nu+q^\mu n^\nu)+\frac{2}{s}q^\mu q^\nu.
\end{equation}
This additional structure is absent at tree level, modifies the angular-momentum distribution and satisfies both
source Ward identities in the distributional sense. 

\begin{figure}
  \makebox[0.48\textwidth][c]{\includegraphics[width=0.45\textwidth,height=203.913bp,keepaspectratio]{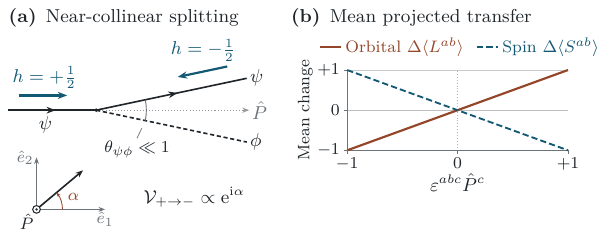}}
  \caption{Spin-to-orbital angular momentum transfer in the real-emission
    channel $\psi\to\psi\phi$. (a) The fermion helicity flips from
    $+\tfrac12$ to $-\tfrac12$. At fixed small opening angle, the phase
    $\mathrm{e}^{\mathrm{i}\alpha}$ winds once as the splitting plane
    rotates about the parent direction $\hat P$, giving the outgoing pair
    one unit of relative orbital angular momentum along $\hat P$.
    (b) The solid and dashed lines show the changes in the mean orbital and
    spin components during splitting. After azimuthal averaging,
    $\Delta\langle L^{ab}\rangle=-\Delta\langle S^{ab}\rangle
    =\hbar\varepsilon^{abc}\hat P^c$ at leading collinear order.}
\label{fig:Splitting}
\end{figure}

If the $\Sigma$-dependent terms in Eq.~\eqref{eq:M-Dirac} are assigned to spin and the remainder to orbital angular momentum, the two contributions contain $1/\epsilon$ poles that cancel in their sum (Appendix~\ref{app:spin}).
In collinear $\phi$ emission, the (anti)fermion flips helicity, and the relative orbital angular momentum of the outgoing pair compensates for the change in spin (Fig.~\ref{fig:Splitting}).
This transfer is reminiscent of a Dirac electron in a central potential, where $\bm{J}=\bm{L}+\bm{S}$ is conserved but $\bm{L}$ and $\bm{S}$ are not separately conserved~\cite{Dirac1928}.

The sum rule~\eqref{eq:conservation} is expressed through the two-current
cut, which factorizes as
$\Pi^{\mu\nu}=\Pi(s)H^{\mu\nu}\Delta_{qq'}$. In four dimensions
$\Pi^{(0)}(s)=e^2/(6\pi)$. We evaluate this cut independently with
AmpRed\cite{chen2025semiautomaticcalculationsmultiloopfeynman} for the
integral reduction and FeynCalc\cite{Shtabovenko:2023idz} for the
algebraic manipulations, finding
\begin{equation}
  \frac{\Pi(s)}{\Pi^{(0)}(s)}
  = 1-\frac{3g_{c}^{2}}{32\pi^{2}}+O(g_{c}^{4}),
  \label{eq:sumrule}
\end{equation}
in agreement with Eq.~\eqref{eq:conservation}.

Since we here adopt the Belinfante definition as the physical angular-momentum flux, the local flux projected to the radial direction is intrinsically by definition Eq.~\eqref{eq:local_I}
\begin{equation}
  n^a \varepsilon^{abc} I^{\mu bc} = r \varepsilon^{abc} \left( n^a n^b \Theta^{\mu c} - n^a n^c \Theta^{\mu b} \right) =0.
\end{equation} 
This should not be regarded to be inconsistent with particles' helicity. Rather, because the angular-momentum detection produces a $\partial \delta$-dependent term, it should be resorted to the wave packet to interpret a physical observable, with a brief discussion based on correlator detecting a free positive-helicity Weyl wave packet in Appendix~\ref{app:helicity}. The helicity of the particle is a global property, while the local angular-momentum flux is a local property, which are not necessarily equal. 

\section{Polarized scattering example\label{sec:example}}

Because the correlators carry a $\partial\delta$ term, the distribution measured in a scattering process is obtained by folding them with the wave packet; we define it as the normalized matrix element
\begin{equation}
  \langle \mathcal{M}^{ab}\rangle_\lambda
  =\frac{\int_{q,q'}\mathcal K^{\lambda}_{\mu\nu}
  \langle\tilde J^{\mu}(q)\mathcal M^{ab}\tilde J^{\nu\dagger}(q')\rangle}
  {\int_{q,q'}\mathcal K^{\lambda}_{\mu\nu}
  \langle\tilde J^{\mu}(q)\tilde J^{\nu\dagger}(q')\rangle},
  \label{eq:Mexp}
\end{equation}
where $\mathcal K^{\lambda}_{\mu\nu}\equiv f(q)f^{*}(q')\varepsilon^{*\lambda}_{\mu}(q)\varepsilon^{\lambda}_{\nu}(q')$ collects the wave packet and the polarizations, $\int_{q,q'}\equiv\int[\mathrm{d} q][\mathrm{d} q']$ with $[\mathrm{d} q]\equiv\mathrm{d}^{D}q/(2\pi)^{D}$ the Lorentz-invariant measure, and $\mathcal M^{ab}$ is evaluated at $\hat n$. 

The distribution in Eq.~\eqref{eq:Mexp} is the expectation value $\operatorname{Tr}[\rho_\lambda\,\mathcal M^{ab}]$ of the angular-momentum detector operator in the normalized state generated by the source $\rho_\lambda=|J_\lambda\rangle\langle J_\lambda|$, where
\begin{equation}
|J_\lambda\rangle=\mathcal N_\lambda^{-1/2}
\int_q f^{*}(q)\varepsilon^{\lambda}_{\nu}(q)
\tilde J^{\nu\dagger}(q)|0\rangle.
\end{equation}
Here, $\mathcal N_\lambda$ is the denominator of Eq.~\eqref{eq:Mexp}. The momentum-space kernel of the density operator $\rho_\lambda$ is $\rho_{Y(q'),X(q)}\equiv\langle Y(q')|\rho_\lambda|X(q)\rangle$. For $q'\neq q$, these matrix elements encode coherence between states with different total momenta. 

A brief view of how the angular-momentum detector contributes differently compared to the energy detector could be demonstrated by multiplying an extra phase factor $\mathrm e^{\mathrm i q_z a_z}$ to the amplitude $f(q)$, which is interpreted as a displacement of the wave packet in the $z$ direction in the real space. The expectation value of the energy detector $\langle \mathcal E \rangle$, defined in Eq.~\eqref{eq:lightray-M}, is insensitive to this phase, while this extra phase directly alters the expectation value of the azimuthal component of the angular-momentum detector, $\mathcal M_\varphi = \cos\varphi\mathcal M^{31} -\sin\varphi\mathcal M^{23}$. The difference in the azimuthal component between the displaced and undisplaced packets is $\langle \mathcal M_\varphi \rangle_{a_z} - \langle \mathcal M_\varphi\rangle_{0} = a_z\sin\theta\langle\mathcal E\rangle $; classically, it is the variation of orbital angular momentum of the centre of mass caused by the displacement of the reference point.

A more interesting choice of wave packet might be to consider a rotating phase around the $z$ axis, which renders orbital angular momentum from a non-relativistic view. To illustrate this, we take a virtual photon produced in a hard scattering process, approximately at rest, with a narrow momentum-space wave packet localized near its central momentum:
\begin{equation}
  \small
  \begin{aligned}
    f_m(q)=\mathcal A_\sigma q_\perp \mathrm{e}^{-\mathrm i m \varphi_q } \exp\left[-\frac{(q^0-Q)^2+q_z^2+q_\perp^2}{4\sigma^2}\right].
  \end{aligned}
  \label{eq:wave_packet}
\end{equation}
Here, $\mathcal A_\sigma$ normalizes the packet, $Q$ is its central energy and $\sigma$ its width, where $\sigma \ll Q$. The variables $q_\perp\equiv|\bm q_\perp|$ and $\varphi_q$ denote the magnitude and azimuthal angle of the transverse momentum, respectively. The azimuthal phase $\mathrm e^{-\mathrm i m\varphi_q}$, with $m=0,\pm1$, selects the orbital-angular-momentum mode about the $z$-axis. All three packets share the same transverse envelope. 

\begin{figure}
  \makebox[0.45\textwidth][c]{\includegraphics[width=0.45\textwidth,height=203.913bp,keepaspectratio]{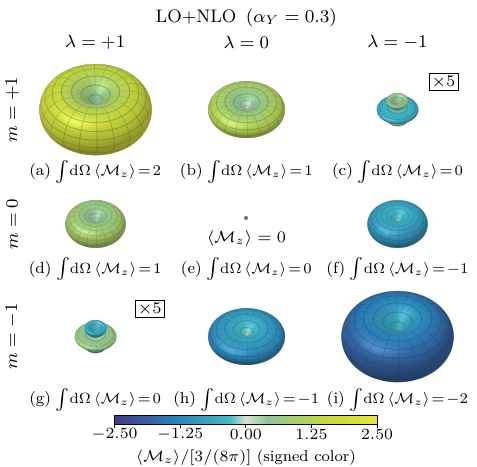}}
  \caption{LO+NLO Belinfante angular-momentum density $\langle\mathcal M_z\rangle$ for narrow packets centered at $q=(Q,\bm 0)$, with $\alpha_Y=g_c^2/(16\pi^2)=0.3$. Rows correspond to $m=+1,0,-1$ and columns to $\lambda=+1,0,-1$. Radius and color encode the absolute and signed density in units of $3/(8\pi)$. Panels (c) and (g), with $(m,\lambda)=(+1,-1)$ and $(-1,+1)$, have vanishing total angular momentum, with a small but nonzero local density; their plotted radii and color values are multiplied by 5.} 
\label{fig:Mz-NLO}
\end{figure}

The detector $\mathcal M^{12} \equiv \mathcal M_z$ measures the $z$ component of angular-momentum flux. In the limit $\sigma/Q \to 0$, with orbital mode $m$ and polarization $\lambda$, the distribution through NLO is 
\begin{equation}
  \begin{aligned}
  &\bigl\langle \mathcal M_z(\hat{n}) \bigr\rangle_{m,\lambda}
  = \frac{3}{8\pi}\sin^2\theta\,\Bigl[m+\lambda \\
  &{}+ \frac{2-3\alpha_Y}{8}\,m(3\lambda^2-2)(5\cos^2\theta-1)
  \Bigr] + O(\alpha_Y^2),
  \end{aligned}
\end{equation}
and Fig.~\ref{fig:Mz-NLO} 
shows the LO+NLO correlators for various packets and polarizations, with
$\alpha_Y=g_c^2/(16\pi^2)=0.3$. Every distribution
integrates to $m+\lambda$, yet their celestial profiles differ,
and the loop correction reshapes some of them without altering the
integrated charge, consistent with Eq.~\eqref{eq:conservation}.
Among the nine combinations of $m$ and $\lambda$, a comparison worth notice is between those with $m+\lambda=0$, which have vanishing total angular momentum but various local distributions, implying that the local angular-momentum displays more detailed information about the scattering. 

In both cases mentioned above, the difference in the phase of the wave packet cannot be detected by the energy flux, but it is resolved by the angular-momentum flux. The phase sensitivity reflects the dependence on first momentum derivatives of the density-matrix kernel $\rho_{Y(q'),X(q)}$ evaluated at $q'=q$, which is dependent on the wave packet.

\section{Conclusion and Discussion}

In this work, we constructed angular-momentum detector operators for massless scalar and Dirac fields and compared the canonical and Belinfante definitions of the local flux. The two definitions yield different Dirac detector operators because the moment arm retains the subleading momentum flux at null infinity. The detector correlator, with two insertions of the conserved vector current, contains derivatives of the momentum-conserving delta function in addition to the usual delta-function term. In massless Yukawa theory, the Belinfante correlator is infrared finite at NLO. In contrast, the angular-momentum detector of canonical definition and spin detector from Belinfante's definition are not infrared-safe in this theory. With the wave-packet examples, we demonstrate that the derivative term makes the angular-momentum flux sensitive to wave-packet structures including momentum-dependent phases invisible to energy detectors. For a related discussion of generalized energy detector we refer to \cite{Korchemsky:2021okt}.

Our perturbative calculation shows that loop corrections modify the distributional structure of the Belinfante angular-momentum flux, while the infrared poles cancel in the correlator considered here. By contrast, at fixed detector direction the canonical correlator retains a one-loop collinear pole whose residue is shown in Appendix~\ref{app:canonical}. This difference demonstrates that definitions of angular-momentum flux that yield the same integrated conserved charge can lead to inequivalent local observables and can exhibit different infrared properties. The cancellation found here makes the Belinfante definition a promising starting point for constructing infrared-finite angular-momentum flux observables. It would be interesting to investigate whether this infrared advantage persists in more general interacting theories and in gravitational settings, where angular-momentum flux has also been studied~\cite{Gonzo_2021}. It would also be interesting to apply our formalism to other physically interesting scattering processes involving angular momentum transfer, such as spin polarization effects in heavy ion collisions~\cite{Liang:2004ph,Liang:2004xn,Fu:2021pok,STAR:2017ckg,STAR:2022fan}.

Integrated over energy, the flux depends only on the direction of observation and thus resolves the angular structure of the outgoing state. In this sense, different detector observables can access different aspects of the momentum-space density matrix. The energy flux is sensitive only to its diagonal components $\rho_{X(q),X(q)}$, whereas the derivative-of-delta-function terms derived above show that the angular-momentum flux operator $\mathcal M^{ab}$ probes first momentum derivatives of the density-matrix kernel at coincident total momenta.  This observation suggests that angular-momentum flux measurements may provide a complementary probe of quantum information in scattering states. Furthermore, higher coordinate moments of light-ray operators~\cite{Cordova:2018ygx,Belin_2021} may probe higher derivatives of the density-matrix kernel and may provide ingredients for a partial reconstruction of the underlying quantum state, in the spirit of quantum-state tomography~\cite{DAriano:2003} and decoherence, which is under active studies in collider physics community recently, see e.g.~\cite{Afik:2020onf,Severi:2021cnj,Cheng:2024rxi,Barr:2024djo,Gu:2025ijz}. Whether a sufficiently complete set of such detector observables can achieve informational completeness is an interesting question for the future.

\acknowledgments

We thank Zhicheng Yang, Shanming Ruan, Haotian Qi for useful discussion.
This work was supported by National Natural Science Foundation of China (NSFC) 
under Grant No. 12425505. HXZ would like to thank the Erwin Schrödinger International 
Institute for Mathematics and Physics (ESI) for hospitality during the program 
``New Paradigms for Harnessing Quantum Field Theory at Colliders”, where part of this 
work was completed. We used LLM for assisting the writing of the paper.

\bibliography{references}

\appendix

\section{Dirac angular-momentum detector in \(D\)-dimensional Yukawa theory\label{app:subleading}}

We derive Eq.~\eqref{eq:M-Dirac} on the detector sphere \(S^{D-2}\).
At \(\bm x=r\hat n\), the area element and the moment arm in
Eq.~\eqref{eq:lightray-M} give a factor \(r^{D-1}\) multiplying
\(n^i(n^a\Theta_D^{ib}-n^b\Theta_D^{ia})\).
Write \(\alpha^c=\gamma^0\gamma^c\), \(\alpha^r=n^c\alpha^c\), and
\(K^{ab}=\kappa^{ab}{}_{c}\alpha^c=n^a\alpha^b-n^b\alpha^a\).
Since \(n^a\partial^b-n^b\partial^a=-\mathrm i\Lambda^{ab}/r\),
the symmetric Dirac stress tensor gives
\begin{equation}
\begin{aligned}
 &n^i(n^a\Theta_D^{ib}-n^b\Theta_D^{ia})\\
 &\quad=\frac{1}{4r}\psi^\dagger\alpha^r
                     \overleftrightarrow{\Lambda}^{ab}\psi
       -\frac{\mathrm i}{4}\psi^\dagger K^{ab}
                     \overleftrightarrow{\partial_r}\psi .
\end{aligned}
\end{equation}
Normal ordering is understood. The two-sided derivatives act only on
the fields, and \(\partial_r\) is taken at fixed \(t\).

At fixed \(u=t-r\), the outgoing radiation field
\(\Psi=\lim_{r\to\infty}r^{(D-2)/2}\psi\) obeys \(\alpha^r\Psi=\Psi\).
Since \(\{\alpha^r,K^{ab}\}=0\), the nominal \(r^{-(D-2)}\) radial
flux vanishes between two leading fields. To extract the surviving
term, we use the Dirac equation at finite \(r\). This order of operations
keeps the distinct radial powers explicit. The angular derivative
\(r^{-1}\nabla_c\psi\) scales as \(r^{-D/2}\), and the Yukawa
source \(\phi\psi\) scales as \(r^{-(D-2)}\). The powers coincide at \(D=4\).
With the coupling convention of Sec.~\ref{sec:oneloop}, the equation
\((\mathrm i\gamma^\mu\partial_\mu+g_c\mu_R^\epsilon\phi)\psi=0\) gives
\begin{equation}
 \partial_r\psi=-\alpha^r\partial_t\psi
 -\frac{1}{r}\alpha^r\alpha^d\nabla_d\psi
 +\mathrm i g_c\mu_R^\epsilon\phi\,\alpha^r\gamma^0\psi .
\end{equation}
Substituting this equation and its adjoint into the radial bilinear gives
the exact identity
\begin{equation}
\begin{aligned}
 \psi^\dagger K^{ab}\overleftrightarrow{\partial_r}\psi
 &=-\partial_t(\psi^\dagger K^{ab}\alpha^r\psi)\\
 &\quad-\frac{1}{r}\bigl[\psi^\dagger K^{ab}\alpha^r
             \alpha^d\nabla_d\psi-\mathrm{h.c.}\bigr].
\end{aligned}
\end{equation}
Indeed, \(\mathrm i\alpha^r\gamma^0\) is Hermitian and commutes with
\(K^{ab}\), since both \(\alpha^r\) and \(\gamma^0\) anticommute
with \(K^{ab}\). The two explicit Yukawa terms therefore cancel at
finite \(r\). We evaluate matrix elements in localized outgoing packets
for which \(r^{D-1}[\psi^\dagger K^{ab}\alpha^r\psi]_{t=0}^{+\infty}\to0\)
and the radiation limit of the fields and their angular derivatives can
be taken under the retarded-time integral.

Both remaining terms have an explicit \(1/r\), so their
\(r^{-(D-1)}\) coefficients are determined by the leading radiation
field. The first term of the stress tensor gives
\(\tfrac14\Psi^\dagger\overleftrightarrow{\Lambda}^{ab}\Psi\).
For the radial contribution use
\(\Psi^\dagger K^{ab}\alpha^r=-\Psi^\dagger K^{ab}\) and
\(\alpha^c\alpha^d=\delta^{cd}+2\mathrm i\Sigma^{cd}\). Its coefficient is
\begin{equation}
\begin{aligned}
 &-\frac{\mathrm i}{4}
    \bigl[\Psi^\dagger K^{ab}\alpha^d\nabla_d\Psi-\mathrm{h.c.}\bigr]\\
 &\quad=\frac14\Psi^\dagger\overleftrightarrow{\Lambda}^{ab}\Psi
  -\frac12\kappa^{ab}{}_{c}\nabla_d
                  (\Psi^\dagger\Sigma^{dc}\Psi).
\end{aligned}
\end{equation}
The metric part of the Clifford product supplies the second quarter
of the orbital term. Its antisymmetric part differentiates the spin
density. Adding the two flux contributions and integrating over \(u\) yields
\begin{equation}
\begin{aligned}
 \mathcal M_D^{ab}(\hat n)
 &=\frac12\int_{-\infty}^{+\infty}\mathrm du\,\Bigl(
     :\Psi^\dagger\overleftrightarrow{\Lambda}^{ab}\Psi:\\
 &\quad-n^a\nabla_c:\Psi^\dagger\Sigma^{cb}\Psi:
       +n^b\nabla_c:\Psi^\dagger\Sigma^{ca}\Psi:\Bigr).
\end{aligned}
\label{eq:M-Dirac-app}
\end{equation}

\section{Canonical Angular-Momentum Detector and Correlator\label{app:canonical}}

For Dirac fields, the canonical angular-momentum flux is the sum of the
orbital and spin contributions, without symmetrizing the energy-momentum
tensor:
\begin{equation}
  \begin{aligned}
    \mathcal{M}_{C}^{ab}(\hat n)
    &= \int\mathrm{d}u\, \left[\frac12:\Psi^{\dagger}
    \overleftrightarrow{\Lambda}^{ab}\Psi:
    +:\Psi^\dagger \Sigma^{ab} \Psi:\right],
    \label{eq:MC}
  \end{aligned}
\end{equation}
This expression follows from the Hermitian Dirac Lagrangian; in
plane-wave modes it becomes
\begin{equation}
  \begin{aligned}
    \mathcal{M}_C^{ab}(\hat n)
    &=\frac{1}{4(2\pi)^{D-1}}
    \int_0^{+\infty}\mathrm{d}\omega\,\omega^{D-3}\\
    &\quad \times\sum_s\Bigl[
    a_{\omega\hat n}^{s\dagger}\overleftrightarrow{\Lambda}^{ab}a_{\omega\hat n}^s
    +b_{\omega\hat n}^{s\dagger}\overleftrightarrow{\Lambda}^{ab}b_{\omega\hat n}^s
    \Bigr],
  \end{aligned}
\label{eq:M-canonical-modes}
\end{equation}

The correlators built from $\mathcal M_D^{ab}$ of
Eq.~\eqref{eq:M-Dirac-modes} and $\mathcal M_C^{ab}$ of
Eq.~\eqref{eq:M-canonical-modes} we denote by
$C_{B}^{ab;\mu\nu}$ and $C_{C}^{ab;\mu\nu}$. 
At the LO, 
\begin{equation}
  C_C^{(0)ab;\mu\nu}
  = F_C^{(0)ab;\mu\nu}\,\Delta_{qq'}
  + G_C^{(0)\mu\nu}\,\mathcal D^{ab}\Delta_{qq'},
  \label{eq:C0-canonical}
\end{equation}
where
\begin{equation}
  \begin{aligned}
    &F_C^{(0)ab;\mu\nu}
    = \frac{\mathrm{i}e^2\mu_R^{2\epsilon}N_D s^{D-2}}
           {2^D(2\pi)^{D-2}w^D}\,\biggl[\\
    &\quad -A^{ab}\Bigl((D-1)w\,P^{\mu\nu}
                  + \frac{s}{w}n^\mu n^\nu\Bigr)\\
    &\quad + \frac{s}{2}W_+^{ab\mu\nu}(n,n)
           - \frac{s}{2}W_-^{ab\mu\nu}(n,n)\\
    &\quad + w\,W_-^{ab\mu\nu}(q,n)
           - w^2\bigl(g^{a\mu}g^{b\nu}
                     -g^{a\nu}g^{b\mu}\bigr)\biggr],
  \end{aligned}
  \label{eq:F0-canonical}
\end{equation}
\begin{equation}
  G_C^{(0)\mu\nu}
  = -e^2\mu_R^{2\epsilon}
    \frac{N_D s^{D-1}}{2^D(2\pi)^{D-2}w^{D-1}}\,
    P^{\mu\nu}.
  \label{eq:G0-canonical}
\end{equation}
At fixed direction
their difference is a total derivative on the celestial sphere, so its
integral vanishes at the LO, while at the NLO the canonical
correlator retains the collinear pole evaluated at $D=4$:
\begin{equation}
  \left.C_C^{(1)ab;\mu\nu}\right|_{\mathrm{pole}}
  =-\frac{g_c^2}{16\pi^2\epsilon}
   \left(C_B^{(0)ab;\mu\nu}-C_C^{(0)ab;\mu\nu}\right).
  \label{eq:can-pole-prl}
\end{equation}
The pole is absent from the integrated charge because the sphere integral
of $C_B-C_C$ vanishes order by order: the two definitions agree for the
total angular momentum but differ as local fluxes.

\section{Angular-Momentum Detector in Spherical-Wave Basis\label{app:sperical}}

In the spherical-wave basis, which diagonalizes total angular-momentum,
the scalar mode operators are related to the plane-wave operators by
\begin{equation}
\begin{gathered}
 c_{\omega lm}
 =\mathrm{i}^{l}\int\frac{\mathrm{d}\Omega_{\hat k}}{4\pi^2}
 Y_{lm}^{*}(\hat k)c_{\omega\hat k},\\
 [c_{\omega' l'm'},c_{\omega lm}^{\dagger}]
  =\frac{\delta(\omega-\omega')\delta_{ll'}\delta_{mm'}}{\pi\omega}.
\end{gathered}
 \label{eq:spw-op-prl}
\end{equation}
For rotations about the $z$ axis, the direction-resolved energy and
angular-momentum fluxes are
\begin{equation}
  \begin{aligned}
    \mathcal E_S(\hat n)
    &=\pi\int\omega^2\mathrm{d}\omega
    \sum_{ll'mm'}\mathrm{i}^{l-l'}\\
    &\quad\times
    Y_{lm}^{*}(\hat n)Y_{l'm'}(\hat n)
    c_{\omega lm}^{\dagger}c_{\omega l'm'},\\
    \mathcal M^{12}_S(\hat n)
    &=\pi\int\omega\mathrm{d}\omega
    \sum_{ll'mm'}\frac{m+m'}{2}\,\mathrm{i}^{l-l'}\\
    &\quad\times
    Y_{lm}^{*}(\hat n)Y_{l'm'}(\hat n)
    c_{\omega lm}^{\dagger}c_{\omega l'm'}.
  \end{aligned}
  \label{eq:spw-scalar}
\end{equation}
Spherical-harmonic orthogonality gives integrated energy and
angular-momentum eigenvalues $\omega$ and $m$; at fixed direction
the operators retain off-diagonal angular modes. 

For massless Dirac fields, spinor spherical harmonics couple orbital angular momentum to spin
$1/2$, and the integration over the full sphere gives eigenvalues
$m$ for particles and $-m$ for antiparticles. In the Weyl basis the
helicity sectors decouple and the Belinfante's angular momentum flux expands separately in
spherical modes, 
\begin{equation}
  \begin{aligned}
    &\mathcal{M}^{12}_{D\pm}(\hat{n}) 
    ={} \pi\int_{0}^{+\infty}\omega\;\mathrm{d}\omega
    \sum_{j,j',m,m'}\mathrm{i}^{j-j'}\\
    &\times\Bigg[
      \frac{m+m'}{2}\,
      \mathcal{W}^{\pm\dagger}_{j,m} \mathcal{W}^{\pm}_{j',m'} a^{\pm\dagger}_{\omega,j,m}a^{\pm}_{\omega,j',m'} \\
      &\;\;\; \mp \frac{1}{4\sin \theta} \partial_\theta \left( \sin^ 2 \theta \, \mathcal{W}^{\pm\dagger}_{j,m} \mathcal{W}^{\pm}_{j',m'} \right)
      a^{\pm\dagger}_{\omega,j,m}a^{\pm}_{\omega,j',m'}\\
    &\;\;\; - \frac{m+m'}{2}\,\mathcal{W}^{\mp\dagger}_{j',m'} \mathcal{W}^{\mp}_{j,m} b^{\pm\dagger}_{\omega,j,m}b^{\pm}_{\omega,j',m'} \\
    &\;\;\; \mp \frac{1}{4\sin \theta} \partial_\theta \left( \sin^ 2 \theta \, \mathcal{W}^{\mp\dagger}_{j',m'} \mathcal{W}^{\mp}_{j,m} \right)b^{\pm\dagger}_{\omega,j,m}b^{\pm}_{\omega,j',m'}
    \Bigg],
  \end{aligned}
  \label{eq:spw-Dirac}
\end{equation}
where $\pm$ denotes the positive and negative
helicity sectors of the angular-momentum detector, the non-zero anti-commutators $\{a^\pm_{\omega' j'm'},a_{\omega jm}^{\pm\dagger}\} =\{b^\pm_{\omega' j'm'},b_{\omega jm}^{\pm\dagger}\} =\frac{\delta(\omega-\omega')\delta_{jj'}\delta_{mm'}}{\pi\omega}$,
and
\begin{equation}
  \mathcal{W}^{\pm}_{j,m} = \frac1{\sqrt2} \left(\mathcal{Y}_{j,j+\frac{1}{2},m}\mp\mathcal{Y}_{j,j-\frac{1}{2},m}\right).
\end{equation}
Here, $\mathcal{Y}_{j,j\pm\frac{1}{2},m}$ denotes the spinor spherical harmonics,
\begin{equation}
  \mathcal Y_{l\pm\frac12,\,l,\,m} =
  \frac{1}{\sqrt{2l+1}}
  \begin{bmatrix}
  \pm\sqrt{l\pm m+\tfrac12}\;Y_{l,\,m-\frac12}\\
  \sqrt{l\mp m+\tfrac12}\;Y_{l,\,m+\frac12}
  \end{bmatrix},
\end{equation}
where $-l\mp\frac12 \leq m \leq l\pm \frac12$ and $Y_{l,\pm(l+1)} \equiv 0$. If the local flux of the canonical definition is used, the angular-momentum detector operator will become
\begin{equation}
  \begin{aligned}
    &\mathcal{M}^{12}_{C\pm}(\hat{n}) 
    ={} \pi\int_{0}^{+\infty}\omega\;\mathrm{d}\omega
    \sum_{j,j',m,m'}\mathrm{i}^{j-j'}\\
    &\qquad\times\Bigg(
      \frac{m+m'}{2}\,
      \mathcal{W}^{\pm\dagger}_{j,m} \mathcal{W}^{\pm}_{j',m'} a^{\pm\dagger}_{\omega,j,m}a^{\pm}_{\omega,j',m'} \\
    &\qquad \quad- \frac{m+m'}{2}\,\mathcal{W}^{\mp\dagger}_{j',m'} \mathcal{W}^{\mp}_{j,m} b^{\pm\dagger}_{\omega,j,m}b^{\pm}_{\omega,j',m'}
    \Bigg).
  \end{aligned}
  \label{eq:spw-Dirac-canonical}
\end{equation}
The Integral of the difference between two detectors over the full sphere is zero. The total angular-momentum charge does not depend on the local flux definitions.

\section{Spin Detector Operator in Yukawa Theory\label{app:spin}}

Picking the terms with the spin matrix $\Sigma$ in Eq.~\eqref{eq:M-Dirac}, we can naively define the so-called spin detector operator in Yukawa theory,
\begin{equation}
  \begin{aligned}
  \mathcal{M}_{\mathrm {spin}}^{ab}(\hat n)
  &= \frac12\int\mathrm{d}u\, \Bigl[
  - n^a \nabla_{c}:\Psi^\dagger \Sigma^{cb} \Psi:\\
  &\quad + n^b \nabla_{c}:\Psi^\dagger \Sigma^{ca} \Psi:\Bigr],
  \end{aligned}\label{eq:M-spin}
\end{equation}
and in mode representation
\begin{equation}
\begin{aligned}
  \mathcal{M}_{\mathrm {spin}}^{ab}(\hat n)
  ={}&\frac{1}{4(2\pi)^{D-1}}
  \int_0^\infty\mathrm{d}\omega\,\omega^{D-3}\\
  &\times
  \Bigl(-n^a \nabla_c \sigma_{\omega\hat n}^{cb}+n^b \nabla_c \sigma_{\omega\hat n}^{ca}\Bigr).
\end{aligned}
\label{eq:M-spin-modes}
\end{equation}
Integrated over the full sphere, the spin-angular-momentum detector read the total spin charge
\begin{equation}
\begin{aligned}
  \int \mathrm d \Omega_{\hat n}\mathcal{M}_{\mathrm {spin}}^{ab}(\hat n)
  ={}\int \frac{\mathrm d^{D-1}{\bm k}}{2k_0(2\pi)^{D-1}} \sigma^{ab}_{\bm k}.
\end{aligned}
\label{eq:total-M-spin-modes}
\end{equation}

Inserting this spin detector into the correlator
\begin{equation}
  \begin{aligned}
    C_{\mathrm{spin}}^{ab;\mu\nu} &\equiv \langle\tilde J^{\mu}(q)\,\mathcal{M}^{ab}_{\mathrm{spin}}(\hat n)\,
    \tilde J^{\nu\dagger}(q')\rangle \\
    &= F^{ab;\mu\nu}_{\mathrm{spin}}\Delta_{qq'} + G_{\mathrm{spin}}^{\mu\nu}\mathcal D^{ab} \Delta_{qq'},
  \end{aligned}
\end{equation}
one can derive the spin distribution before the source normalization up to NLO in Yukawa theory. Since the spherical derivatives over $\hat n$ can be factored out of the loop and phase space integrals, $G_{\mathrm{spin}}=0$ holds to all orders in perturbation theory. The coefficient $F_{\mathrm{spin}}$ to NLO is
\begin{gather}
\begin{aligned}
 &F_{\rm spin}^{(0)ab;\mu\nu}
 =\frac{\mathrm{i}e^2s^2}{16\pi^2w^4}
 \biggl[
 \Bigl(\tfrac32s-wq\cdot \tau\Bigr)W_-^{ab\mu\nu}(n,n)\\
 &\quad-\tfrac32\Bigl(wW_-^{ab\mu\nu}(n,q)+A^{ab}A^{\mu\nu}\Bigr)\\
 &\quad+w^2W_-^{ab\mu\nu}(n,\tau)+wA^{ab}(n^\mu\tau^\nu-n^\nu\tau^\mu)
 \biggr],
\end{aligned}
\label{eq:Bspin-F0}\\
\begin{aligned}
 &F_{\rm spin}^{(1)ab;\mu\nu}
 =\frac{\mathrm{i}e^2g_c^2s^2}{256\pi^4w^4}
 \biggl\{\\
 &\;\;\;\Bigl[\bigl(\tfrac32s-wq\cdot \tau\bigr)Z_{\rm s}-s\Bigr] W_-^{ab\mu\nu}(n,n)\\
 &\;\;\;-\bigl(\tfrac32Z_{\rm s}-1\bigr)
 \Bigl(wW_-^{ab\mu\nu}(n,q)+A^{ab}A^{\mu\nu}\Bigr)\\
 &\;\;\;+Z_{\rm s}\Bigl[
 w^2W_-^{ab\mu\nu}(n,\tau)+wA^{ab}(n^\mu\tau^\nu-n^\nu\tau^\mu)
 \Bigr]\biggr\},
\end{aligned}
\label{eq:Bspin-F1}
\end{gather}
with dimensional regularization, where we define the unit timelike vector $\tau^\mu = (1,\bm 0)$ and the parameter
\begin{equation}
\begin{aligned}
 Z_{\rm s} =\frac{1}{\epsilon}+\frac72-\gamma_E +3\ln(4\pi)
 +\ln\!\left(\frac{\mu_R^4 w^2}{s^3}\right).
\end{aligned}
\label{eq:Bspin-L}
\end{equation}

This spin-orbital separation depends on the observer through \(\tau^\mu\), and the spin contribution retains a collinear pole at one loop. 

\section{Helicity and Belinfante Angular-Momentum Detector\label{app:helicity}}

We illustrate the distinction between helicity and radial angular-momentum flux using a free positive-helicity Weyl particle in four dimensions. Throughout, we retain the spinor prescribed in the main text,
\begin{equation}
  \chi_{\bm q}^{+} = \sqrt{2\omega_{\bm q}} \begin{bmatrix}
    \cos \frac{\theta_{\bm q}}{2} \mathrm e^{-\mathrm i \frac{\varphi_{\bm q}}{2}}, & \sin \frac{\theta_{\bm q}}{2} \mathrm e^{\mathrm i \frac{\varphi_{\bm q}}{2}}
  \end{bmatrix}^T,
\end{equation}
satisfying $\chi_{\bm{q}}^{+\dagger}\chi_{\bm{q}}^+=2\omega_{\bm q}$ and $(\Lambda^{12}+\Sigma^{12})\chi_{\bm{q}}^+=0$. We consider the Gaussian source
\begin{equation}
\label{eq:helicity-gaussian-source}
\begin{aligned}
 f_\sigma(q)&=\mathrm e^{-\mathrm i\varphi_{\bm{q}}/2}
 \exp\!\left[-\frac{(q_z-Q)^2+q_\perp^2}{4\sigma^2}\right],\\
 |\Psi_\sigma\rangle&=\mathcal N_\sigma^{-1/2}
 \int[\mathrm dq]\,f^*_\sigma(q)\tilde\psi^\dagger(q)\chi_{\bm{q}}^+|0\rangle,
 \end{aligned}
\end{equation}
with $Q>0$, $\sigma\ll Q$, and normalization factor $\mathcal N_\sigma = \int [\mathrm d q] [\mathrm d q'] f_\sigma(q)\chi_{\bm q}^{+\dagger} \langle \tilde{\psi}(q) \tilde{\psi}^\dagger(q')\rangle f^*_\sigma(q')\chi_{\bm q'}^{+}$. Only its values on $q^0=\omega$ enter the free state. Here, an extra phase factor is introduced to ensure that the full projecting amplitude is basically well-defined as a function of the angular coordinates except the South pole,
\begin{equation}
 f_\sigma^{*}(q) \chi^+_{\bm q} \propto \begin{bmatrix}
  \cos \frac{\theta_{\bm{q}}}2, &
  \sin \frac{\theta_{\bm{q}}}2\,\mathrm e^{\mathrm i\varphi_{\bm{q}}}
 \end{bmatrix}^T.
\end{equation}
Also, this phase is chosen to ensure the projecting amplitude degenerates to a spinor $\propto \begin{bmatrix} 1,\ 0 \end{bmatrix}^T$ when $\theta_{\bm{q}} = 0$, which represents the mode along $z$-axis without an additional vortex phase.

The Belinfante detector expectation is defined by 
\begin{equation}
\langle \mathcal M_{z}(\hat n) \rangle = \langle\Psi_\sigma|\mathcal M^{12}(\hat n)|\Psi_\sigma\rangle.
\end{equation} 
Let $p(\hat n)\,\mathrm d\Omega_{\hat{n}}$ denote the probability that the particle momentum points into $\mathrm d\Omega_{\hat{n}}$, i.e.
\begin{equation}
\begin{aligned}
p(\hat n)&=\frac{1}{\mathcal N_\sigma}
 \int \frac{|\bm q|^2\mathrm d |\bm q|\mathrm d q^0}{(2\pi)^4}\,[\mathrm d q']\,
 f_\sigma(q)f^*_\sigma(q')\\
 &\quad\times \chi_{\bm q}^{+\dagger}
 \bigl\langle \tilde{\psi}(q)\tilde{\psi}^\dagger(q')\bigr\rangle
 \chi_{\bm q'}^{+}.
\end{aligned}
\end{equation}

Free contractions of the Belinfante operator give
\begin{equation}
\label{eq:belinfante-angular-density}
\begin{aligned}
\langle \mathcal M_{z}(\hat n) \rangle
 &=\frac{1}{2}p
   -\frac{1}{4\sin\theta}
    \partial_\theta(\sin^2\theta\,p)\\
 &=\frac{1-\cos\theta}{2}p
   -\frac{\sin\theta}{4}\partial_\theta p .
 \end{aligned}
\end{equation}
Consequently,
\begin{equation}
\int \mathrm d\Omega_{\hat{n}}\,\langle \mathcal M_{z}(\hat n) \rangle=\frac12,
 \qquad \langle \mathcal M_{z}(\hat n) \rangle\Big|_{\theta=0}=0.
\end{equation}
If we hold $Q>0$ fixed and take $\sigma\to0^+$, in the forward region $\theta=O(\sigma/Q)$, the leading profiles are
\begin{equation}
\begin{aligned}
 p(\hat n)&\simeq
 \frac{Q^2}{2\pi\sigma^2}
 \mathrm e^{-\frac{Q^2\theta^2}{2\sigma^2}},\\
\langle \mathcal M_{z}(\hat n) \rangle&\simeq
 \frac{Q^4\theta^2}{8\pi\sigma^4}
 \mathrm e^{-\frac{Q^2\theta^2}{2\sigma^2}}.
 \end{aligned}
\end{equation}
At any fixed $\theta \gg \sigma/Q$, the contribution also tends to zero; the nonzero integral is carried by a shrinking angular region around the axis. The detector limit is taken at finite packet width before $\sigma\to0$. This distinguishes the local radial projection $\varepsilon^{abc}n^c{\mathcal M^{ab}}(\hat n)=0$ from the particle's helicity.

\section{Discrete Symmetry in Massless Yukawa Theory\label{app:symmetry}}

In four dimensions massless Yukawa theory requires an independent quartic
coupling for renormalizability,
\begin{equation}
 \mathcal L=
 \frac12(\partial\phi)^2+\frac{\mathrm{i}}2\bar\psi\overleftrightarrow{\slashed{\partial}}\psi
 +g_c\bar\psi\psi\phi+\frac{h_c}{4!}\phi^4,
 \label{eq:Lmassless-prl}
\end{equation}
with $h_c=O(g_c^2)$, which does not enter the correlators here. The
interaction is invariant under
\begin{equation}
 \psi\longrightarrow\gamma^5\psi,\qquad
 \bar\psi\longrightarrow-\bar\psi\gamma^5,\qquad
 \phi\longrightarrow-\phi ,
 \label{eq:Z2-prl}
\end{equation}
while $\bar\psi\psi$ and $\phi^3$ are odd, so fermion masses and a
cubic scalar coupling are forbidden to all orders. Scalar mass and quartic
coupling are even. We set the renormalized scalar mass to zero and let
$h_c$ run freely. The symmetry only classifies counterterms, hence no
$\gamma^5$ insertion is needed in the regulated integrands.
\end{document}